\documentclass[12pt,english,aps,manuscript]{article}
\usepackage[T1]{fontenc}
\usepackage[latin9]{inputenc}
\usepackage{geometry}
\usepackage{amsmath}
\usepackage{amssymb}
\usepackage[numbers]{natbib}
\usepackage{babel}
\begin{document}
\begin{center}
\textbf{\Large{}Radiative Generation of dS from AdS }{\Large\par}
\par\end{center}

\begin{center}
\vspace{0.3cm}
\par\end{center}

\begin{center}
{\large{}S. P. de Alwis$^{\dagger}$ }{\large\par}
\par\end{center}

\begin{center}
Physics Department, University of Colorado, \\
 Boulder, CO 80309 USA 
\par\end{center}

\begin{center}
\vspace{0.3cm}
\par\end{center}
\begin{abstract}
The large volume scenario (LVS) of type IIB string compactifications
has a robust supersymmetry breaking minimum with a negative cosmological
constant (CC). We argue that radiative corrections below the Kaluza-Klein
(KK) scale can result in a positive CC, though the string theory generated
CC is negative, if some mild conditions on the specterum of low energy
fluctuations are satisfied. This would make the so-called deSitter
swampland conjecture (even if true at a high scale) physically irrelevant. 
\end{abstract}
\vfill{}

$^{\dagger}$ dealwiss@colorado.edu 

\eject

\section{Introduction}

The difficulties of getting a classical solution of string theory
that corresponds to a 4D dS space are well-known. Even with the inclusion
of non-perturbative effects in type IIB compactifications on a CYO
as in \citep{Kachru:2003aw} (KKLT), and \citep{Balasubramanian:2005zx}
(LVS), unless one adds additional structure (such as Dbar branes,
T branes etc.), which makes these constructions more complicated and
less well-controlled, the background (vacuum) solutions are AdS.

That said, there is an important difference between the (pre-uplift)
KKLT and the LVS vacua. While the former preserves (${\cal N}$=1)
supersymmetry, the latter breaks it. We will therefore focus on the
AdS vacuum of LVS. The point of this note is that subject to some
mild conditions discussed below, the SUSY breaking LVS minimum with
a negative CC can be lifted to a positive CC, by radiative corrections
coming from the fluctuations of states below the compactification
scale\footnote{For a related discussion see \citep{deAlwis:2019aud}.}.

\section{Review of LVS}

The Kaehler and superpotential for the moduli sector (setting $M_{P}=1$)
are

\begin{eqnarray}
\hat{K} & = & -2\ln\left({\cal V}+\frac{\xi}{2}\left(\frac{(S+\bar{S})}{2}\right)^{3/2}\right)-\ln\left(i\int\Omega\wedge\bar{\Omega}(U,\bar{U})\right)-\ln(S+\bar{S}),\label{eq:hatK}\\
\hat{W} & = & \int G_{3}\wedge\Omega+\sum_{i}A_{i}e^{-a_{i}T^{i}}\equiv W_{0}+W_{{\rm np}}.\label{eq:hatW}
\end{eqnarray}
$\xi$= $-\chi\zeta(3)/2(2\pi)^{3}$, $\chi$ is the Euler character
of the CYO $S$ the axidilaton, $U$ the complex structure moduli,
${\cal V}$ the volume of the CYO (taken to be of the ``Swiss cheese''
type) in Planck units, $T^{i}$ are Kaehler moduli whose real parts
determine 4-cycle volumes. $W_{0}$ is the flux superpotential and
$W_{{\rm np}}$ is that generated by non-perturbative effects. The
$a_{i}$ a constants determined by string instantons\footnote{These come from Euclidean D3 branes wrapping a 4 cycle \citep{Witten:1996bn}
in which case $a=2\pi$.} or from confinement/gaugino condensation effects - for a pure $SU(n)$
gauge theory on D7 branes $a=8\pi^{2}/n$. In the simplest Swiss cheese
constructions ${\cal V}=k_{b}\tau_{b}^{3/2}-k_{s}\tau_{s}^{3/2}$
where $\tau_{b}=\Re T_{b}$ is a big cycle determining the volume
and $\tau_{s}=\Re T_{s}$ is a small cycle $k_{b,s}$ are order one
numbers. In a realistic model we would have more than this minimum
number of 4 cycles and furthermore there would be additional terms
in $K,$ and $W$ coming from the visible matter (MSSM/SGUT) sector.
The latter play no role in determining the potential for the moduli
and the background values of these fields are effectively zero at
this stage.

The axi-dilaton $S$ and the complex structure moduli are fixed super-symmetrically
in terms of the internal fluxes i.e. as solutions of $D_{S}W=0,\,D_{T^{i}}W=0$.
Then one gets the following potential for the $T_{i}$,

\begin{eqnarray}
V & = & V_{F}+V_{D}\label{eq:pot1}\\
V_{F} & = & \frac{4}{3}g_{s}(a|A|)^{2}\frac{\sqrt{\tau_{s}}e^{-2a\tau_{s}}}{{\cal V}}-2g_{s}a|AW_{0}|\frac{\tau_{s}e^{-a\tau_{s}}}{{\cal V}^{2}}+\frac{3}{8}\frac{\xi|W_{0}|^{2}}{g_{s}^{1/2}{\cal V}^{3}}+\ldots\label{eq:VF}\\
V_{D} & = & \frac{f}{2}D^{2},\,D=f^{-1}k^{i}K_{i}\label{eq:VD}
\end{eqnarray}
The ellipses in the second equation represent higher powers of $1/{\cal V}$
and the expansion is justified for large volume compactifications,
which will be obtained by appropriate choice of fluxes. The minimization
conditions give, 
\begin{eqnarray}
e^{-a\tau_{s}} & \simeq & \frac{3}{4}\frac{W_{0}}{aA{\cal V}}\sqrt{\tau_{s}}\left(1-\frac{3}{4a\tau_{s}}\right),\label{eq:sol1}\\
\tau_{s}^{3/2} & \simeq & \frac{\hat{\xi}}{2}(1+\frac{1}{2a\tau_{s}})\simeq\frac{\hat{\xi}}{2},\label{eq:sol2}
\end{eqnarray}
where we've written $\hat{\xi}=(\frac{S+\bar{S}}{2})^{3/2}\xi=\xi/g_{{\rm s}}^{3/2}$.
Note that extremizing with respect to $\tau_{s}$ gives us an exponentially
large volume and the three displayed terms in $V_{F}$ are all of
order ${\cal V}^{-3}$. This would mean that that at the classical
(negative) minimum found in \citep{Balasubramanian:2005zx}, $V_{D}=0$
since it is positive definite and of order $1/{\cal V}^{2}$. This
would also be the case with the contribution to the F-term potential
from the dilaton and complex-structure moduli.

The minimum found in \citep{Balasubramanian:2005zx} gives a negative
value for the (classical) cosmological constant
\begin{equation}
V_{0}\simeq-\frac{3\hat{\xi}}{16a\tau_{s}}\frac{m_{3/2}^{2}}{{\cal V}}\simeq-\frac{m_{3/2}^{2}}{\ln m_{3/2}}\frac{1}{{\cal V}},\label{eq:V0}
\end{equation}
where $m_{3/2}$ is the mass of the gravitino. Since the F-terms of
the Kaehler moduli are non-zero at this minimum this solution breaks
supersymmetry spontaneously.

\section{Radiative corrections to the CC}

Now we wish to estimate the radiative corrections due to quantum fluctuations
around this minimum. These would be due to the moduli fluctuations
around the background values fixed by the LVS minimum as well as the
supersymmetric standard model and (super) GUT states (if present). 

The RG improved one-loop contribution (in a spontaneously broken SUSY
theory) to the cosmological constant evaluated in the deep IR (i.e.
at cosmological scales) due to these states is \footnote{The original Coleman-Weinberg formula \citep{Coleman:1973jx} was
adapted for SUSY/SUGRA theories in \citep{Ferrara:1994kg}. The version
below which incorporates one-loop RG running from the KK scale is
taken from \citep{DeAlwis:2021gou}.}

\begin{align}
\Lambda_{{\rm cc}}(\mu & \ll m)\simeq\Lambda_{{\rm cc}}(M_{{\rm KK}})-\frac{\bar{M}^{2}-M_{{\rm KK}}^{2}}{32\pi^{2}}{\rm Str}_{\bar{M}}{\bf M}^{2}-\frac{\bar{m}^{2}-\bar{M}^{2}}{32\pi^{2}}{\rm Str}_{\bar{m}}{\bf m^{2}}\nonumber \\
 & +\frac{1}{64\pi^{2}}{\rm Str}_{\bar{M}}{\bf M}^{4}\ln\left(\frac{\bar{M}^{2}}{M_{{\rm KK}}^{2}}\right)+\frac{1}{64\pi^{2}}{\rm Str}_{\bar{m}}{\bf m}^{4}\ln\left(\frac{\bar{m}^{2}}{\bar{M}^{2}}\right)+\ldots.\label{eq:CC0susy}
\end{align}
The cutoff of the effective field theory for these fluctuations is
the Kaluza-Klein scale $M_{KK}=\frac{O(1)}{{\cal V}^{2/3}}$. We've
split up the states below this scale into two sectors, one characterized
by the scale $\bar{M}$ refers to all the heavy moduli as well as
matter supermultiplets at some high scale such as the GUT scale. The
other set would contain the MSSM states as well as the light moduli.
We've assumed for convenience that the splitting within a supermultiplet
is smaller than the separation between these two scales.

The RHS of this equation should be identified with the observed CC,
i.e. $\Lambda_{{\rm cc}}(\mu\ll m)\simeq10^{-122}.$ The first term
on the LHS however is the cosmological constant derived derived in
LVS, i.e. we identify it with the LVS minimum \eqref{eq:V0}, so (using
also \eqref{eq:sol2}),
\begin{equation}
\Lambda_{{\rm cc}}(M_{{\rm KK}})=V_{0}\sim-\frac{3\hat{\xi}}{16a\tau_{s}}\frac{m_{3/2}^{2}}{{\cal V}}=-\frac{3}{16a}\frac{(4\xi)^{1/3}}{\sqrt{g_{{\rm s}}}}\frac{m_{3/2}^{2}}{{\cal V}}.\label{eq:CCKK}
\end{equation}
The CC at this scale is of course negative. The question is whether
the radiative corrections can do the job of lifting this (SUSY broken)
AdS vacuum to the observed value.

The validity of the EFT implies that the largest mass scale allowed
is much less than the cutoff scale i.e. $\bar{M}\ll M_{KK}$. Also
supertraces are essentially fixed by the gravitino mass and the number
of states at or below this scale. In fact we have (see for example
\citep{Bagger:1990qh} and \citep{Ferrara:2016ntj} for the correction
when the background is not Minkowski\footnote{I wish to thank Heliudson de Oliveira Bernardo for pointing out this
reference to me. })

\begin{equation}
\frac{1}{2}{\rm Str}{\bf M}^{2}=(N-1)(m_{3/2}^{2}+V_{0})-F^{I}({\cal R}_{I\bar{J}}+S_{I\bar{J}})F^{\bar{J}},\label{eq:Str}
\end{equation}
where $N$ is the number of supermultiplets contributing to the supertrace,
${\cal R}_{I\bar{J}}=\partial_{I}\partial_{J}\ln\det K_{M\bar{N}}$
and $S_{I\bar{J}}=-\partial_{I}\partial_{\bar{J}}\ln\det\Re f_{ab}$
with $K_{M\bar{N}}$ being the matter sector Kaehler metric and $f_{ab}$
the gauge coupling superfield. The sums over $I,J$ effectively go
only over the SUSY breaking directions whose dimension is typically
a number of order one (in our case it is essentially just the $T_{b}$
direction with a small contribution from $T_{s}$). In other words
it does not scale with $N$. Hence for $N\gg1$ we have, since $V_{0}$
is suppressed by an extra factor of ${\cal V}$ compared to $m_{3/2}^{2}$
in LVS (unlike in KKLT where $m_{3/2}^{2}+V_{0}=-2m_{3/2}^{2}$),
\begin{equation}
{\rm Str}{\bf M}^{2}\simeq Nm_{3/2}^{2}.\label{eq:Strapprox}
\end{equation}
 Thus for large $N$ it follows from \eqref{eq:CC0susy} that the
radiative contributions are positive and thus we have the possibility
of lifting the negative CC coming from the string theory calculation
to a positive value. 

Let us consider whether such a scenario is actually plausible - starting
from the LVS minimum of eqn. \eqref{eq:CC0susy}. Keeping just the
dominant terms in \eqref{eq:CC0susy} we get.
\begin{align}
\Lambda_{{\rm cc}}(\mu & \ll m)\simeq\Lambda_{{\rm cc}}(M_{{\rm KK}})+\frac{M_{{\rm KK}}^{2}-\bar{M}^{2}}{32\pi^{2}}{\rm Str}_{\bar{M}}{\bf M}^{2}+\frac{\bar{M}^{2}-\bar{m}^{2}}{32\pi^{2}}{\rm Str}_{\bar{m}}{\bf m^{2}}\nonumber \\
 & \simeq\Lambda_{{\rm cc}}(M_{{\rm KK}})+\frac{M_{{\rm KK}}^{2}}{16\pi^{2}}Nm_{3/2}^{2}\equiv V_{0}+\Delta_{{\rm radiative}}V.\label{eq:DV}
\end{align}
Here we've used $M_{KK}\gg\bar{M}$ and eqn. \eqref{eq:Strapprox}.
To have a chance of cancelling the negative LVS CC with the radiative
corrections we should have $\Delta_{{\rm radiative}}V\gtrsim|V_{0}|$
which gives (using the estimate $M_{KK}^{2}\simeq1/{\cal V}^{4/3}$
and \eqref{eq:CCKK}) an upper bound
\begin{equation}
{\cal V}^{1/3}\lesssim\left(\frac{\sqrt{g_{{\rm s}}}}{(4\xi)^{1/3}}\right)\left(\frac{N}{16\pi^{2}}\right)\left(\frac{16a}{3}\right).\label{eq:volbound}
\end{equation}
 For typical LVS compactifications $\chi\sim10^{2}$ and the number
of heavy moduli (the complex structure moduli) are also $O(10^{2})$,
and if we have a SGUT we can have an additional number of heavy states
(i.e. with mass greater than the electro-weak scale but much less
than the KK scale), so that one expects $N\sim10^{2}$. So the first
product of the first two factora can easily be a number of the order
unity. The third factor can however be large. Taking for instance
$a=2\pi$ which is the value generated by D3 brane instantons (as
in \citep{Witten:1996bn}) it is about 30 so we get an upper bound
\[
{\cal V}\lesssim10^{3}.
\]
 Note that one can get larger upper-bounds for smaller Euler characters
(keeping $g_{{\rm s}}<1$) and larger $N$ as well as from larger
values of $a$ coming from gaugino condensation. Now one might think
that the species bound 
\[
M_{KK}=\frac{M_{P}}{{\cal V}^{2/3}}\lesssim\frac{M_{P}}{\sqrt{N}},
\]
gives a constraint. However this translates to (using \eqref{eq:volbound}),
\[
\sqrt{N}\lesssim{\cal V}^{2/3}\lesssim\left(\frac{\sqrt{g_{{\rm s}}}}{(4\xi)^{1/3}}\right)^{2}\left(\frac{N}{16\pi^{2}}\right)^{2}\left(\frac{16a}{3}\right)^{2},
\]
which just gives a lower bound on $N$. For the choices of $a=2\pi$
and $\chi=10^{2}$, this gives $N\gtrsim O(10)$.

Of course the mere fact that we get a radiatively generated dS vacuum
in the IR, does not necessarily mean that we get the observed value
of the CC. For that we need to use the Bousso-Polchinski \citep{Bousso:2000xa}
argument for getting the right CC by scanning over the landscape of
LVS flux vacua, and in that respect we have added nothing new to the
CC problem. The point of this note is to show that even if one can
rigorously prove that it is impossible to get a positive CC at the
KK scale, it does not mean that the the CC in the IR cannot be positive.
In fact our argument shows that provided that the states below the
KK scale satisfy some mild criteria, one can generate a positive CC
in the IR.

\section{Acknowledgement}

I wish to thank Fernando Quevedo for discussions.

\bibliographystyle{apsrev}
\bibliography{myrefs}

\end{document}